# A unified approach to mapping and clustering of bibliometric networks

Ludo Waltman, Nees Jan van Eck, and Ed C.M. Noyons

Centre for Science and Technology Studies, Leiden University, The Netherlands
{waltmanlr, ecknjpvan, noyons}@cwts.leidenuniv.nl

In the analysis of bibliometric networks, researchers often use mapping and clustering techniques in a combined fashion. Typically, however, mapping and clustering techniques that are used together rely on very different ideas and assumptions. We propose a unified approach to mapping and clustering of bibliometric networks. We show that the VOS mapping technique and a weighted and parameterized variant of modularity-based clustering can both be derived from the same underlying principle. We illustrate our proposed approach by producing a combined mapping and clustering of the most frequently cited publications that appeared in the field of information science in the period 1999–2008.

## 1. Introduction

In bibliometric and scientometric research, a lot of attention is paid to the analysis of networks of, for example, documents, keywords, authors, or journals. Mapping and clustering techniques are frequently used to study such networks. The aim of these techniques is to provide insight into the structure of a network. The techniques are used to address questions such as:

- What are the main topics or the main research fields within a certain scientific domain?
- How do these topics or these fields relate to each other?
- How has a certain scientific domain developed over time?

To satisfactorily answer such questions, mapping and clustering techniques are often used in a combined fashion. Various different approaches are possible. One approach is to construct a map in which the individual nodes in a network are shown and to display a clustering of the nodes on top of the map, for example by marking off areas in the map that correspond with clusters (e.g., McCain, 1990; White & Griffith, 1981) or by coloring nodes based on the cluster to which they belong (e.g., Leydesdorff & Rafols, 2009; Van Eck, Waltman, Dekker, & Van den Berg, 2010). Another approach is to first cluster the nodes in a network and to then construct a map in which clusters of nodes are shown. This approach is for example taken in the work of Small and colleagues (e.g., Small, Sweeney, & Greenlee, 1985) and in earlier work of our own institute (e.g., Noyons, Moed, & Van Raan, 1999).

In the bibliometric and scientometric literature, the most commonly used combination of a mapping and a clustering technique is the combination of multidimensional scaling and hierarchical clustering (for early examples, see McCain, 1990; Peters & Van Raan, 1993; Small et al., 1985; White & Griffith, 1981). However, various alternatives to multidimensional scaling and hierarchical clustering have been introduced in the literature, especially in more recent work, and these alternatives are also often used in a combined fashion. A popular alternative to multidimensional scaling is the mapping technique of Kamada and Kawai (1989; see e.g. Leydesdorff & Rafols, 2009; Noyons & Calero-Medina, 2009), which is sometimes used together with the pathfinder network technique (Schvaneveldt, Dearholt, & Durso, 1988; see e.g. Chen, 1999; de Moya-Anegón et al., 2007; White, 2003). Two other alternatives to multidimensional scaling are the VxOrd mapping technique (e.g., Boyack, Klavans, & Börner, 2005) and our own VOS mapping technique (e.g., Van Eck et al., 2010). Factor analysis, which has been used in a large number of studies (e.g., de Moya-Anegón et al., 2007; Leydesdorff & Rafols, 2009; Zhao & Strotmann, 2008), may be seen as a kind of clustering technique and, consequently, as an alternative to hierarchical clustering. Another alternative to hierarchical clustering is clustering based on the modularity function of Newman and Girvan (2004; see e.g. Wallace, Gingras, & Duhon, 2009; Zhang, Liu, Janssens, Liang, & Glänzel, 2010).

As we have discussed, mapping and clustering techniques have a similar objective, namely to provide insight into the structure of a network, and the two types of techniques are often used together in bibliometric and scientometric analyses. However, despite their close relatedness, mapping and clustering techniques have typically been developed separately from each other. This has resulted in techniques that have little in common. That is, mapping and clustering techniques are based on different ideas and rely on different assumptions. In our view, when a mapping and a clustering technique are used together in the same analysis, it is generally desirable that the techniques are based on similar principles as much as possible. This enhances the transparency of the analysis and helps to avoid unnecessary technical complexity. Moreover, by using techniques that rely on similar principles, inconsistencies between the results produced by the techniques can be avoided. In this paper, we propose a unified approach to mapping and clustering of bibliometric networks. We show how a mapping and a clustering technique can both be derived from the same underlying principle. In doing so, we establish a relation between on the one hand the VOS mapping technique (Van Eck & Waltman, 2007; Van Eck et al., 2010) and on the other hand clustering based on a weighted and parameterized variant of the well-known modularity function of Newman and Girvan (2004).

The paper is organized as follows. We first present our proposal for a unified approach to mapping and clustering. We then discuss how the proposed approach is related to earlier work published in the physics literature. Next, we illustrate an application of the proposed approach by producing a combined mapping and clustering of frequently cited publications in the field of information science. Finally, we summarize the conclusions of our research. Some technical issues are elaborated in appendices.

## 2. Mapping and clustering: A unified approach

Consider a network of $n$ nodes. Suppose we want to create a mapping or a clustering of these nodes. $c_{ij}$ denotes the number of links (e.g., co-occurrence links, co-citation links, or bibliographic coupling links) between nodes $i$ and $j$ ($c_{ij} = c_{ji} \geq 0$). $s_{ij}$ denotes the association strength of nodes $i$ and $j$ (Van Eck & Waltman, 2009) and is given by

$$s_{ij} = \frac{2mc_{ij}}{c_i c_j}, \tag{1}$$

where $c_i$ denotes the total number of links of node $i$ and $m$ denotes the total number of links in the network, that is,

$$c_i = \sum_{j \neq i} c_{ij} \qquad \text{and} \qquad m = \frac{1}{2}\sum_i c_i \, . \tag{2}$$

In the case of mapping, we need to find for each node $i$ a vector $x_i \in \mathrm{R}^p$ that indicates the location of node $i$ in a $p$-dimensional map (usually $p = 2$). In the case of clustering, we need to find for each node $i$ a positive integer $x_i$ that indicates the cluster to which node $i$ belongs. Our unified approach to mapping and clustering is based on minimizing

$$V(x_1, \ldots, x_n) = \sum_{i<j} s_{ij} d_{ij}^2 - \sum_{i<j} d_{ij} \tag{3}$$

with respect to $x_1, \ldots, x_n$. $d_{ij}$ denotes the distance between nodes $i$ and $j$ and is given by

$$d_{ij} = \left\| x_i - x_j \right\| = \sqrt{\sum_{k=1}^{p} (x_{ik} - x_{jk})^2} \tag{4}$$

in the case of mapping and by

$$d_{ij} = \begin{cases} 0 & \text{if } x_i = x_j \\ 1/\gamma & \text{if } x_i \neq x_j \end{cases} \tag{5}$$

in the case of clustering. We refer to the parameter $\gamma$ in (5) as the resolution parameter ($\gamma > 0$). The larger the value of this parameter, the larger the number of clusters that we obtain. Equation (3) can be interpreted in terms of attractive and repulsive forces between nodes. The first term in (3) represents an attractive force, and the second term represents a repulsive force. The higher the association strength of two nodes, the stronger the attractive force between the nodes. Since the strength of the repulsive force between two nodes does not depend on the association strength of the nodes, the overall effect of the two forces is that nodes with a high association strength are pulled towards each other while nodes with a low association strength are pushed away from each other.

In the case of mapping, it has been shown that the above approach is equivalent to the VOS mapping technique (Van Eck & Waltman, 2007; Van Eck et al., 2010), which is in turn closely related to the well-known technique of multidimensional scaling.

In the case of clustering, it can be shown (see Appendix A) that minimizing (3) is equivalent to maximizing

$$\hat{V}(x_1, \ldots, x_n) = \frac{1}{2m} \sum_{i<j} \delta(x_i, x_j) w_{ij} \left( c_{ij} - \gamma \frac{c_i c_j}{2m} \right), \tag{6}$$

where the weights $w_{ij}$ are given by

$$w_{ij} = \frac{2m}{c_i c_j} \, . \tag{7}$$

Interestingly, if the resolution parameter $\gamma$ and the weights $w_{ij}$ are set equal to 1 in (6), then (6) reduces to the so-called modularity function introduced by Newman and Girvan (2004; see also Newman, 2004b). Clustering (also referred to as community detection) based on this modularity function (Newman, 2004a) is very popular among physicists and network scientists (for an extensive overview of the literature, see Fortunato, 2010). In bibliometric and scientometric research, modularity-based clustering has been used in a number of recent studies (Lambiotte & Panzarasa, 2009; Schubert & Soós, 2010; Takeda & Kajikawa, 2009; Wallace et al., 2009; Zhang et al., 2010). It follows from (6) and (7) that our proposed clustering technique can be seen as a kind of weighted variant of modularity-based clustering (see Appendix B for a further discussion). However, unlike modularity-based clustering, our clustering technique has a resolution parameter $\gamma$. This parameter helps to deal with the resolution limit problem (Fortunato & Barthélemy, 2007) of modularity-based clustering. Due to this problem, modularity-based clustering may fail to identify small clusters. Using our clustering technique, small clusters can always be identified by choosing a sufficiently large value for the resolution parameter $\gamma$.

## 3. Related work

Our unified approach to mapping and clustering is related to earlier work published in the physics literature. Here we summarize the most closely related work.

The above result showing how mapping and clustering can be performed in a unified and consistent way resembles to some extent a result derived by Noack (2009). Noack defined a parameterized objective function for a class of mapping techniques (referred to as force-directed layout techniques by Noack). This class of mapping techniques includes for example the well-known technique of Fruchterman and Reingold (1991). Noack showed that his parameterized objective function subsumes the modularity function of Newman and Girvan (2004). In this way, Noack established a relation between on the one hand a class of mapping techniques and on the other hand modularity-based clustering. Our result differs from the result of Noack in three ways. First, the result of Noack does not directly relate well-known mapping techniques such as the one of Fruchterman and Reingold to modularity-based clustering. Instead, Noack's result shows that the objective functions of some well-known mapping techniques and the modularity function of Newman and Girvan are special cases of the same parameterized function. Our result establishes a direct relation between a mapping technique that has been used in various applications, namely the VOS mapping technique, and a clustering technique. Second, the mapping and clustering techniques considered by Noack and the ones that we consider differ from each other by a weighing factor. This is the weighing factor given by (7). Third, the clustering technique considered by Noack is unparameterized, while our clustering technique has a resolution parameter $\gamma$.

A parameterized variant of the modularity function of Newman and Girvan (2004) was introduced by Reichardt and Bornholdt (2006; see also Heimo, Kumpula, Kaski, & Saramäki, 2008; Kumpula, Saramäki, Kaski, & Kertész, 2007). Clustering based on this generalized modularity function is closely related to our proposed clustering technique. In fact, setting the weights $w_{ij}$ equal to 1 in (6) essentially yields the function of Reichardt and Bornholdt.

## 4. Illustration of the proposed approach

We now illustrate an application of our unified approach to mapping and clustering. In Figure 1, we show a combined mapping and clustering of the 1242 most frequently cited publications that appeared in the field of information science in the period 1999–2008.[1] The mapping and the clustering were produced using our unified approach. This was done as follows. We first collected an initial set of publications. This set consisted of all Web of Science publications of the document types article and review published in 37 information science journals in the period 1999–2008 (for the list of journals, see Van Eck et al., 2010, Table 1). Publications without references were not included. We then extended the initial set of publications with all Web of Science publications in the period 1999–2008 cited by or referring to at least five publications in the initial set of publications. In this way, we ended up with a set of 9948 publications. For each publication in this set, we counted the number of citations from other publications in the set. We selected the 1242 publications with at least eight citations for further analysis. For these publications, we determined the number of co-citation links and the number of bibliographic coupling links. These two types of links were added together and served as input for both our mapping technique and our clustering technique.[2] In the case of our clustering technique, we tried out a number of different values for the resolution parameter $\gamma$. After some experimenting, we decided to set this parameter equal to 2. This turned out to yield a clustering with a satisfactory level of detail.

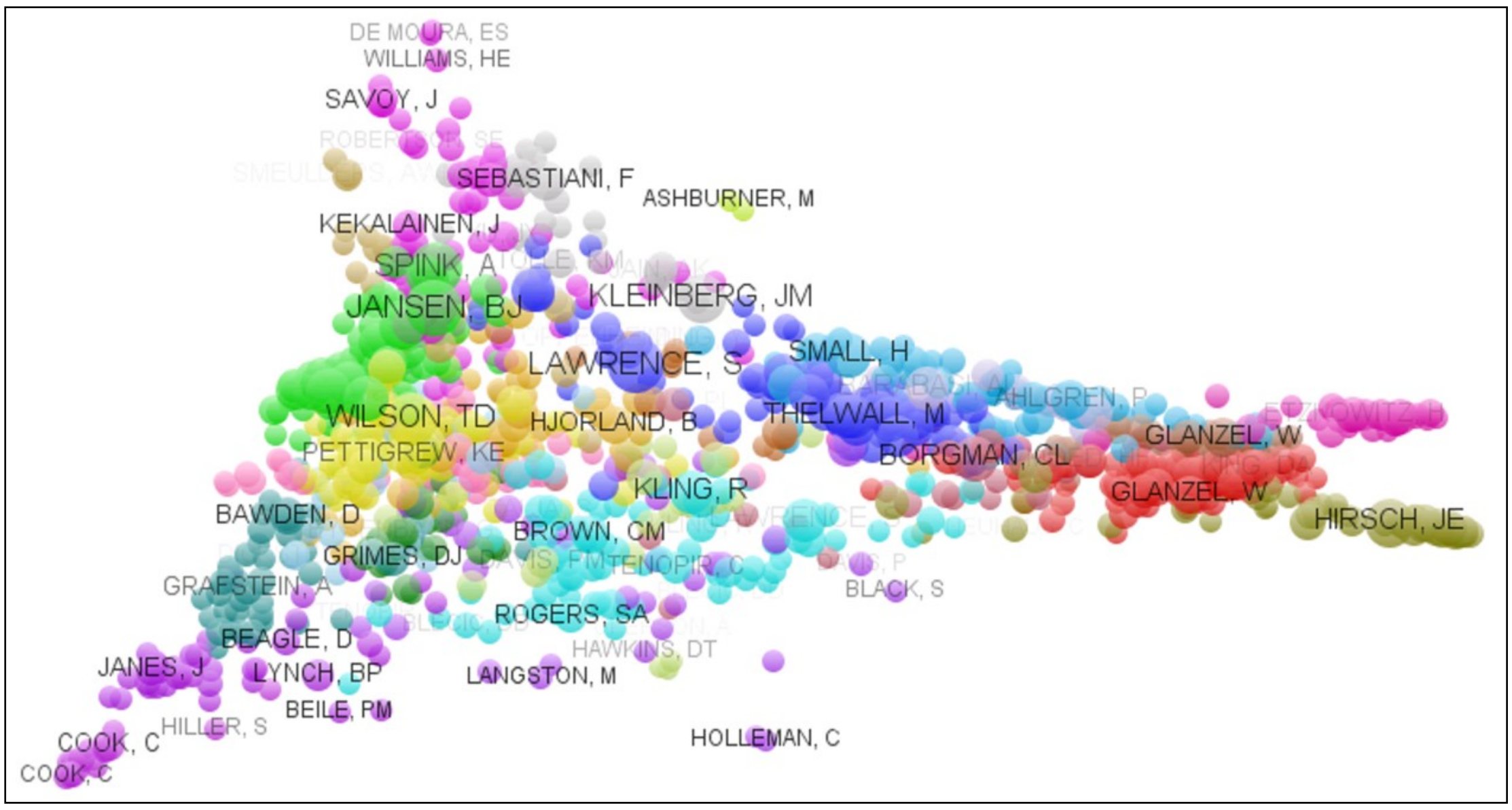

Figure 1. Combined mapping and clustering of the 1242 most frequently cited publications that appeared in the field of information science in the period 1999–2008. Publications are labeled with the name of the first author. Colors are used to indicate clusters.

---

[1] For other bibliometric studies of the field of information science at the level of individual publications, we refer to Åström (2007) and Chen, Ibekwe-SanJuan, and Hou (in press).

[2] Our techniques for mapping and clustering both require solving an optimization problem. In the case of mapping, we minimized (3) using a majorization algorithm (similar to Borg & Groenen, 2005, Chapter 8). In the case of clustering, we maximized (8) using a top-down divisive algorithm combined with some local search heuristics.

The combined mapping and clustering shown in Figure 1 provides an overview of the structure of the field of information science. The left part of the map represents what is sometimes referred to as the information seeking and retrieval (ISR) subfield (Åström, 2007), and the right part of the map represents the informetrics subfield. The distinction between these two subfields is well known and has been observed in a number of studies (e.g., the influential study of White & McCain, 1998). Within the ISR subfield, a further distinction can be made between "hard" (system-oriented) and "soft" (user-oriented) research (e.g., Åström, 2007). Hard ISR research is located in a relatively small area in the upper left part of our map, while soft ISR research is located in a much larger area in the middle and lower left part of the map.

The clustering shown in Figure 1 consists of 25 clusters. The distribution of the number of publications per cluster has a mean of 49.7 and a standard deviation of 31.5. There is one very small cluster consisting of just two publications. These two publications are concerned with the use of information science techniques to support biological research. The largest cluster consists of 123 publications. The publications in this cluster deal with citation analysis and some related bibliometric and scientometric topics. Out of the 25 clusters, eight clusters are used to cover the informetrics subfield. We have examined these clusters in more detail. A summary of the contents of the eight informetrics clusters is provided in Table 1.

Table 1. Summary of the contents of the eight informetrics clusters. The four authors with the largest number of publications in a cluster are listed as important authors.

| No of pub. | Important authors | Main topics |
|---|---|---|
| 123 | Rousseau, R.; Glänzel, W.; Moed, H.F.; Van Raan, A.F.J. | Citation analysis; research evaluation; general scientometric topics |
| 101 | Thelwall, M.; Vaughan, L.; Bar-Ilan, J.; Wilkinson, D. | Webometrics |
| 73 | Leydesdorff, L.; Chen, C.M.; White, H.D.; Small, H. | Mapping and visualization of science |
| 53 | Egghe, L.; Burrell, Q.L.; Daniel, H.D.; Glänzel, W. | *h*-index; citation distributions; Google Scholar |
| 48 | Glänzel, W.; Cronin, B.; Bozeman, B.; Shaw, D. | Scientific collaboration; co-authorship |
| 46 | Meyer, M.; Leydesdorff, L.; Tijssen, R.J.W.; Zimmermann, E. | Science and technology studies; patent analysis |
| 26 | Nisonger, T.E.; Cronin, B.; Shaw, D.; Wilson, C.S. | Studies of the library and information science field |
| 14 | Newman, M.E.J.; Barabasi, A.L.; Albert, R.; Jeong, H. | Complex networks; scientific collaboration networks |

The results presented above illustrate an application of our unified approach to mapping and clustering. Our approach seems to yield an accurate and detailed picture of the structure of the field of information science. The interested reader is invited to examine the results in more detail at www.ludowaltman.nl/unified_approach/. On this web page, the combined mapping and clustering shown in Figure 1 can be inspected using the VOSviewer software (Van Eck & Waltman, in press). The clustering is also available in a spreadsheet file.

## 5. Conclusions

Mapping and clustering are complementary to each other. Mapping can be used to obtain a fairly detailed picture of the structure of a bibliometric network. For practical purposes, however, the picture will usually be restricted to just two dimensions.

Hence, relations in more than two dimensions will usually not be visible. Clustering, on the other hand, does not suffer from dimensional restrictions. However, the price to be paid is that clustering works with binary rather than continuous dimensions. As a consequence, clustering tends to provide a rather coarse picture of the structure of a bibliometric network.[3]

Given the complementary nature of mapping and clustering and given the frequent combined use of mapping and clustering techniques, we believe that a unified approach to mapping and clustering can be highly valuable. A unified approach ensures that the mapping and clustering techniques on which one relies are based on similar ideas and similar assumptions. By taking a unified approach, inconsistencies between the results produced by mapping and clustering techniques can be avoided.

In this paper, we have elaborated a proposal for a unified approach to mapping and clustering. Our proposal unifies the VOS mapping technique with a weighted and parameterized variant of modularity-based clustering. As discussed elsewhere (Van Eck & Waltman, 2007; Van Eck et al., 2010), the VOS mapping technique is closely related to the well-known technique of multidimensional scaling, which has a long history in the statistical literature (for an extensive overview, see Borg & Groenen, 2005). Modularity-based clustering, on the other hand, is a recent result from the physics literature (Newman, 2004a, 2004b; Newman & Girvan, 2004). It follows from this that our proposed unified approach establishes a connection between on the one hand a long-lasting research stream in the field of statistics and on the other hand a much more recent research stream in the field of physics.

Our unified approach to mapping and clustering can be especially useful when multiple maps of the same domain are needed, each at a different level of detail. For example, when bibliometric mapping is used for science policy purposes, two maps may be needed. On the one hand a detailed map may be needed that can be carefully validated by experts in the domain of interest, and on the other hand a much more general map may be needed that can be provided to science politicians and research managers. The former map may show the individual nodes in a bibliometric network, while the latter map may show clusters of nodes. Expert validation, which is a crucial step in the use of bibliometric mapping for science policy purposes (Noyons, 1999), of course only makes sense when the map presented to domain experts shows essentially the same structure of the domain of interest as the map presented to science politicians. A unified approach to mapping and clustering helps to avoid discrepancies between maps constructed at different levels of detail. In that way, a unified approach facilitates the use of bibliometric mapping in a science policy context.

In the latest version of our freely available VOSviewer software (Van Eck & Waltman, in press; see www.vosviewer.com), we have incorporated algorithms that implement our unified approach to mapping and clustering. Open source algorithms to be run in MATLAB are available at www.ludowaltman.nl/unified_approach/.

[3] In this paper, we have been concerned with clustering techniques that require each node in a bibliometric network to be assigned to exactly one cluster. These are the most commonly used clustering techniques. We have not discussed clustering techniques that allow nodes to be assigned to multiple clusters (e.g., Fortunato, 2010, Section 11). The latter techniques provide a more detailed picture of the structure of a bibliometric network.

## Appendix A

In this appendix, we prove that in the case of clustering minimizing (3) is equivalent to maximizing (6) with weights $w_{ij}$ given by (7). Using (1) and (5), it can be seen that (3) can be rewritten as

$$V(x_1,\ldots,x_n)=\frac{1}{\gamma}\sum_{i<j}(1-\delta(x_i,x_j))\left(\frac{1}{\gamma}\frac{2mc_{ij}}{c_i c_j}-1\right), \tag{8}$$

where $\delta(x_i, x_j)$ equals 1 if $x_i = x_j$ and 0 otherwise. Let us define

$$\hat{V}(x_1,\ldots,x_n)=-\frac{\gamma^2}{2m}V(x_1,\ldots,x_n)+\frac{1}{2m}\sum_{i<j}\left(\frac{2mc_{ij}}{c_i c_j}-\gamma\right). \tag{9}$$

Notice that (9) is obtained from (8) by multiplying with a constant and by adding a constant. The multiplicative constant is always negative. It follows from this that minimizing (8) is equivalent to maximizing (9). Substituting (8) into (9) yields

$$\hat{V}(x_1,\ldots,x_n)=\frac{1}{2m}\sum_{i<j}\delta(x_i,x_j)\left(\frac{2mc_{ij}}{c_i c_j}-\gamma\right). \tag{10}$$

We have now shown that minimizing (3) is equivalent to maximizing (10). Furthermore, (10) can be rewritten as (6) with weights $w_{ij}$ given by (7). This completes the proof.

## Appendix B

Our proposed clustering technique can be seen as a weighted and parameterized variant of modularity-based clustering. Modularity-based clustering maximizes (6) with weights $w_{ij}$ that are set equal to 1. Our clustering technique maximizes (6) with weights $w_{ij}$ that are given by (7). In this appendix, we provide an illustration of the effect of the weights $w_{ij}$ in (7).

Consider a network of $n = 31$ nodes. Let

$$c_{ij}=\begin{cases}10 & \text{if } 1\le i\le 10 \text{ and } 1\le j\le 10 \text{ and } i\ne j\\ 100 & \text{if } 11\le i\le 20 \text{ and } 11\le j\le 20 \text{ and } i\ne j\\ 100 & \text{if } 21\le i\le 30 \text{ and } 21\le j\le 30 \text{ and } i\ne j\\ 20 & \text{if } (1\le i\le 10 \text{ and } j=31) \text{ or } (i=31 \text{ and } 1\le j\le 10)\\ 50 & \text{if } (11\le i\le 20 \text{ and } j=31) \text{ or } (i=31 \text{ and } 11\le j\le 20)\\ 0 & \text{otherwise.}\end{cases} \tag{11}$$

Our clustering technique (with the resolution parameter $\gamma$ set equal to 1) and modularity-based clustering both identify three clusters. They both produce a cluster that contains nodes 1, …, 10, another cluster that contains nodes 11, …, 20, and a third cluster that contains nodes 21, …, 30. However, the two clustering techniques do not agree on the cluster to which node 31 should be assigned. Our clustering technique assigns node 31 to the same cluster as nodes 1, …, 10, while modularity-based clustering assigns node 31 to the same cluster as nodes 11, …, 20. The disagreement on the assignment of node 31 is due to the effect of the weights $w_{ij}$ in (7). It follows from (7) that, compared with modularity-based clustering, our clustering technique gives less weight to nodes with a larger total number of links. Nodes 11, …, 20 have a much larger total number of links than nodes 1, …, 10, and compared with modularity-based clustering our clustering technique therefore gives less weight to nodes 11, …, 20 and more weight to nodes 1, …, 10. Node 31 is strongly associated both with nodes 1, …, 10 and with nodes 11, …, 20. However, due to the difference in weighting, our clustering technique assigns node 31 to the same cluster as nodes 1, …, 10 while modularity-based clustering assigns node 31 to the same cluster as nodes 11, …, 20.

Which of the two assignments of node 31 is to be preferred? The total number of links of nodes 11, …, 20 is almost an order of magnitude larger than the total number of links of nodes 1, …, 10, but the number of links between node 31 and nodes 11, …, 20 is only 2.5 times larger than the number of links between node 31 and nodes 1, …, 10. Hence, from a relative point of view, node 31 has more links with nodes 1, …, 10 than with nodes 11, …, 20. Based on this observation, assigning node 31 to the same cluster as nodes 1, …, 10 seems preferable to assigning node 31 to the same cluster as nodes 11, …, 20. Hence, we believe that, at least in this particular example, the results produced by our clustering technique are preferable to the results produced by modularity-based clustering.